\documentclass[12pt]{article}
\usepackage{epsf}
\usepackage{amsmath, amssymb}

\usepackage[dvips]{graphicx}

\usepackage{comment}

\begin{document}

\newcommand{\lsim}{\stackrel{<}{_\sim}}
\newcommand{\gsim}{\stackrel{>}{_\sim}}

\newcommand{\rem}[1]{{$\spadesuit$\bf #1$\spadesuit$}}

\renewcommand{\theequation}{\thesection.\arabic{equation}}

\renewcommand{\thefootnote}{\fnsymbol{footnote}}
\setcounter{footnote}{0}

\begin{titlepage}

\def\thefootnote{\fnsymbol{footnote}}

\begin{center}

\hfill UT-13-04\\

\vskip .75in

{\Large \bf 

  Non-Thermal Production of Wino Dark Matter \\
  via the Decay of Long-Lived Particles

}

\vskip .75in

{\large
Takeo Moroi, Minoru Nagai and Masahiro Takimoto
}

\vskip 0.25in

{\em Department of Physics, University of Tokyo,
Tokyo 113-0033, Japan}

\end{center}
\vskip .5in

\begin{abstract}

  We consider supersymmetric models in which the neutral Wino is the
  lightest superparticle (LSP), and study the possibility that
  non-thermally produced Wino plays the role of dark matter.  The
  thermal relic density of Wino is smaller than the present mass
  density of dark matter if $m_{\tilde{W}}\lesssim 2.9\ {\rm TeV}$;
  however, even with smaller Wino mass, the Wino can be the dominant
  component of dark matter if it is non-thermally produced by the
  decay of a long-lived particle.  In order to study such a
  possibility in detail, we perform a precise calculation of the
  present mass density of Wino produced by the decay of a long-lived
  particle, taking account of the following effects: (i)
  coannihilation among charged and neutral Winos, and (ii) Sommerfeld
  effect on the pair annihilation cross section of Winos.  We consider
  several well-motivated cases where the long-lived particle
  corresponds to cosmological moduli fields, gravitino, or axino, and
  discuss the implication of the Wino LSP for these cases.

\end{abstract}

\end{titlepage}

\renewcommand{\thepage}{\arabic{page}}
\setcounter{page}{1}
\renewcommand{\thefootnote}{\#\arabic{footnote}}
\setcounter{footnote}{0}

\section{Introduction}
\label{sec:introduction}
\setcounter{equation}{0}

In models with low-energy supersymmetry (SUSY), there may exist
various long-lived particles (which we call $X$) which have very weak
interactions with the particles in the minimal SUSY standard model
(MSSM).  The examples include moduli fields, Polonyi field
(responsible for SUSY breaking), gravitino (which is the superpartner
of graviton), axino (which is the superpartner of axion in SUSY
Peccei-Quinn model), and so on.  In most of the cases, these particles
are irrelevant for low-energy phenomenology because of the weakness of
their interactions.

Cosmologically, however, they often cause serious problems
\cite{Coughlan:1983ci, Ellis:1986zt, Goncharov:1984qm, Banks:1993en,
  de Carlos:1993jw} because, even though their interactions are very
weak, a sizable amount of them may be produced in the early universe.
If they are stable, they survive until today, resulting in the
overclosure of the universe.  Even if they are unstable, their
lifetimes are often so long that they decay at a very late stage of
the evolution of the universe.  If they decay after the big-bang
nucleosynthesis (BBN) starts, hadro- and photo-dissociation processes
of light elements induced by their decay products may spoil the
success of the BBN scenario.  In order to avoid such a problem, the
lifetime of $X$ is required to be shorter than $\sim 1\ {\rm sec}$;
otherwise, the abundance of $X$ is bounded above
\cite{Kawasaki:2004yh, Kawasaki:2004qu}.

Even with lifetime shorter than $\sim 1\ {\rm sec}$, $X$ may cause
other cosmological difficulties.  In our study, we consider one of
such problems: the overproduction of the lightest superparticle (LSP)
due to its decay.  In many cases, $X$ decays into superparticles which
cascade down to the LSP.  Then, if all the produced LSPs survive until
today, it is often the case that the resultant mass density of the LSP
becomes much larger than the present energy density of dark matter.
In particular, the universe may be once dominated by $X$ before the
decay of $X$; in such a case, the overproduction of the LSP is a
serious problem.

If the pair annihilation cross section of the LSP is large enough, the
LSPs produced by the decay of $X$ may annihilate and the abundance of
the LSP can be suppressed.  There exists a viable and well-motivated
candidate of such LSP, the neutral Wino $\tilde{W}^0$.  In models with
the naive grand-unified-theory (GUT) relation among gaugino masses,
the Wino cannot be the LSP.  However, even if the unification of the
gauge group is realized, the GUT relation can be easily violated in
various cases, for example, in the anomaly-mediation model
\cite{Giudice:1998xp, Randall:1998uk}, product-group unification
\cite{Yanagida:1994vq}, and so on.  In particular, in the light of the
recent discovery of the Higgs-like particle at the LHC
\cite{Aad:2012gk, Chatrchyan:2012gu}, the anomaly-mediation model with
large scalar masses \cite{Giudice:1998xp} (as large as $\sim 10-100\
{\rm TeV}$) is well-motivated because the relatively large SUSY Higgs
mass of $\sim 126\ {\rm GeV}$ can be realized if the scalar tops are
so heavy \cite{Giudice:2011cg}.  It is well-known that the Wino can
naturally be the LSP in anomaly-mediation model.

In this paper, we reconsider the possibility of the non-thermally
produced Wino being dark matter.  In particular, we precisely
calculate the relic abundance of the neutral Wino.  The possibility of
Wino dark matter from the decay of long-lived particles was first
discussed in \cite{Giudice:1998xp, Moroi:1999zb}, in which it was
shown that the relic density of the Wino can be consistent with the
dark matter density.  Then, such a scenario has been applied to
various cases \cite{Acharya:2008zi, Acharya:2008bk, Baer:2011hx,
  Baer:2011uz, Moroi:2011ab}.  (For the case where the LSP is not
Wino-like, see also \cite{Gelmini:2006pw, Gelmini:2006pq,
  Nagai:2007ud, Nagai:2008se}.)  Compared to the previous studies, we
have carefully taken into account the following in the calculation of
the relic Wino abundance:
\begin{itemize}
\item[(i)] Coannihilation effect among neutral and charged Winos,
  which becomes important when the decay temperature of $X$ is higher
  than the mass difference between charged and neutral Winos.
\item[(ii)] Sommerfeld enhancement of the pair annihilation cross
  section, which is large when the Wino is heavier than $\sim 1\ {\rm
    TeV}$.
\end{itemize}

The organization of this paper is as follows.  In Section
\ref{sec:formulas}, we show relevant formulas to calculate the relic
abundance of neutral Wino.  In Section \ref{sec:numerical}, we
numerically calculate the relic abundance of the Wino for several
situations and discuss implications.  Section \ref{sec:conclusions} is
devoted to conclusions and discussion.

\section{Formulas}
\label{sec:formulas}

In this section, we summarize the basic formulas to calculate the
thermal relic abundance of the neutral Wino through the decay of a
heavy particle $X$.

The Wino is $SU(2)_L$-triplet, and there exist neutral and charged
Winos, denoted as $\tilde{W}^0$ and $\tilde{W}^\pm$, respectively.  In
the present study, we consider the case where the neutral Wino is the
LSP; in the following, we assume that superparticles other than Winos
are so heavy that they are irrelevant at the time of the freeze-out of
Winos.  This is the case in the anomaly-mediated model, which is one
of the important motivations of our study.  In addition, because we
assume that the Higgsinos are much heavier than Winos, the mass
difference between $\tilde{W}^0$ and $\tilde{W}^\pm$ are dominantly
from one-loop diagrams with gauge bosons (i.e., $\gamma$, $Z$, and
$W^\pm$) and Winos inside the loop.  Then, if the Wino mass is much
smaller than the Higgsino mass, the mass difference is typically
$150-165\ {\rm MeV}$ \cite{Feng:1999fu, Ibe:2012sx}, which is
insensitive to the Wino mass.

Because the mass difference is much smaller than the Wino mass, the
number density of the charged Wino at the time of the freeze-out of
Winos may be sizable.  Thus, in the calculation of the relic
density of $\tilde{W}^0$, we include effects of all the possible
channels of Wino annihilation.  We consider the case where Winos
are much lighter than other superparticles, so the coannihilation
with those are neglected in the calculation of the relic density.

Now we discuss the evolution of the number density of Wino.  As we
mentioned, we consider the case with a long-lived particle $X$ which
decays into MSSM particles.  Using the fact that all the MSSM
superparticles eventually decay into charged or neutral Wino, the
relevant set of Boltzmann equations to calculate the relic abundance
of Wino LSP is given by
\begin{eqnarray}
  \frac{d n_{\tilde{W}}}{d t} + 3 H n_{\tilde{W}}
  &=&  - \langle \sigma_{\rm eff} v \rangle 
  ( n_{\tilde{W}}^2 - n_{\tilde{W}, {\rm eq}}^2)
  + N_{\tilde W} \Gamma_X n_X,
  \label{nwinodot} \\ 
  \frac{d n_X}{d t} + 3 H n_X 
  &=& - \Gamma_X n_X ,
  \label{nxdot} \\ 
  \frac{d \rho_{\rm rad}}{d t} 
  \left(
    1+\frac{1}{3} \frac{\partial \ln g_{\ast}}{\partial \ln T}
  \right) &=& 
\left( -4H{\rho}_{\rm rad}+q \right)  \left( 1+\frac{1}{4}\frac{\partial \ln g_{\ast}}{\partial \ln T}\right),
  \label{rhoraddot}
\end{eqnarray}
where $n_{\tilde{W}}$ is the sum of the number densities of neutral
and charged Winos, $n_X$ is the number density of $X$, and $q$ is a
heat injection into radiation as
\begin{eqnarray}
  q&=&(m_X -  N_{\tilde W} m_{{\tilde W}})\Gamma_X n_X 
  + m_{\tilde W} \langle \sigma_{\rm eff}  v \rangle n_{\tilde{W}}^2,
 \label{heat}
\end{eqnarray}
with $N_{\tilde W}$ being the averaged number of SUSY particles
produced by the decay of one $X$.  In addition, $\rho_{\rm rad}$ is
the energy density of the relativistic component, and is related to
the cosmic temperature $T$ as
\begin{eqnarray}
  \rho_R = \frac{\pi^2}{30} g_*(T) T^4,
\end{eqnarray}
where $g_*(T)$ is the effective number of relativistic degrees of
freedom.\footnote
{We use the fact that $g_{*s}(T)$ is numerically very close to
  $g_{*}(T)$, and approximate $g_{*s}(T)\simeq g_{*}(T)$ in our
  calculation, where $g_{*s}(T)$ is the effective number of massless
  degrees of freedom for the calculation of entropy density, which is
  related to the entropy density as
  \begin{eqnarray*}
    s(T) = \frac{2\pi^2}{45} g_{*s}(T) T^3.
  \end{eqnarray*}
} 
In our calculation, we approximated that the full particle content at
the temperature above the QCD scale (which is taken to be $200\ {\rm
  MeV}$ in our analysis) is that of the MSSM, while that at the
temperature below the QCD scale consists of photon, three generations
of leptons, and pions.  Furthermore, $n_{\tilde{W}, {\rm eq}}$ denotes
the thermal-equilibrium value of $n_{\tilde{W}}$, $H$ is the expansion
rate of the universe, $\Gamma_X$ is the decay rate of $X$, and $m_X$
and $m_{\tilde{W}}$ are the masses of $X$ and Wino,
respectively.\footnote
{Because we are interested in the case where charged and neutral Winos
  are quite degenerate, we denote the Wino masses as $m_{\tilde{W}}$
  as far as we discuss the quantities which are insensitive to the
  mass difference.}
In the above Boltzmann equations, the thermally-averaged effective
annihilation cross section $\langle \sigma_{\rm eff} v \rangle$
accounts both for the coannihilation effect and the Sommerfeld effect,
which were not fully taken into accounts in previous analyses.

The coannihilation processes are included by summing up the cross
sections of all the relevant modes with appropriate weights:
\begin{equation}
 \langle \sigma_{\rm eff} v \rangle = 
 \sum_{i,j} r_i r_j
 \langle \sigma_{ij} v \rangle ,
\end{equation} 
where $i,j = {\tilde W}^0$, ${\tilde W}^+$ and ${\tilde W}^-$, and
\begin{eqnarray}
  r_i = \frac{n_i}{n_{\tilde{W}}},
\end{eqnarray}
with $n_i$ being the number density of $i$.  We assume that Winos are
in kinetic equilibrium;\footnote
{ The energetic Winos produced by the $X$ decay show non-trivial
  velocity distribution at first.  However, charged Winos soon lose
  their energies through the electromagnetic interactions with the
  thermal background (in particular, electron and positron) as well as
  through the decay.  In addition, sizable fractions of neutral Winos
  can be thermalized through the inelastic interactions
  \cite{Hisano:2000dz, Ibe:2012hr}.  In particular, as we see
    below, $T_X$ is required to be higher than $\sim \Delta
    m_{\tilde{W}}$ in order to realize
    $\Omega_{\tilde{{W}}}=\Omega_{\rm c}$ in the parameter region
    where the pair annihilation of the Wino becomes effective; in such
    a case, the neutral and charged Winos are efficiently converted to
    each other by the charged current processes in the thermal bath,
    and the charged Wino efficently loses its energy by the decay
    process (as well as the scattering processes with charged
    particles in the thermal bath).  Thus, we expect that the Winos
    reach the kinetic equilibrium.}
when $T\ll m_{\tilde{W}}$,
\begin{eqnarray}
  r_{\tilde{W}^0} = \frac{1}{1 + 2e^{-\Delta m_{\tilde{W}}/T}},
  ~~~
  r_{\tilde{W}^+} = r_{\tilde{W}^-} = 
  \frac{e^{-\Delta m_{\tilde{W}}/T}}{1 + 2e^{-\Delta m_{\tilde{W}}/T}}.
\end{eqnarray}
If the mass difference between charged and neutral Winos $\Delta
m_{\tilde{W}}$ is much larger than the background temperature $T$,
only the lightest neutral Wino is relevant in the annihilation
process.  On the other hand, the coannihilation becomes effective for
the temperature $T \gtrsim \Delta m_{\tilde{W}}$.

For each annihilation process, thermally averaged cross section is
obtained by
\begin{equation}
  \langle \sigma_{ij} v \rangle = 
  \left( \frac{m_{{\tilde W}}}{4\pi T} \right)^{3/2}
  \int d^3 v  (\sigma_{ij} v)
  e^{-m_{{\tilde W}} v^2/4T}.
\end{equation}
Once the temperature of the universe decreases and the neutral and
charged Winos become non-relativistic, the wave functions of
annihilating Wino pairs are significantly deformed by the electroweak
potential generated by the electroweak gauge boson exchanges.  The
resultant annihilation cross sections, $\sigma_{ij} v$, are
significantly enhanced or suppressed due to the Sommerfeld effect
\cite{Hisano:2004ds, Hisano:2005ec, Hisano:2006nn}.  The Sommerfeld
effect is more important for larger Wino mass, $m_{{\tilde W}}\gsim 1\
{\rm TeV}$, because the electroweak potential behaves as a long range
force in such a mass region.  As a result, the thermally-averaged
cross sections show non-trivial dependence on the cosmic temperature
$T$ and the Wino mass $m_{{\tilde W}}$.  A two-body system of Winos
can be classified by the quantum numbers $Q$ (electric charge) and $S$
(spin), and the Sommerfeld enhancement factors are evaluated for fixed
values of these quantum numbers.  For each set of $(Q, S)$, possible
decay modes and the decay widths are summarized in Appendix.  (See
Table \ref{Tab:cs}.)

\begin{figure}[t]
\label{fig:cs}
  \centerline{\epsfxsize=0.6\textwidth\epsfbox{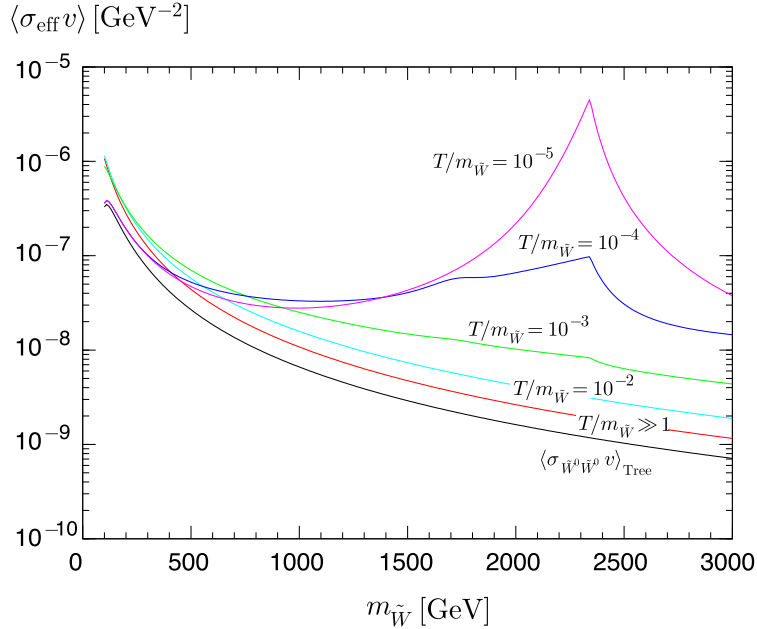}}
  \caption{Thermally averaged total cross section $\langle \sigma_{\rm
      eff} v \rangle$ for $T= 10^{-5}m_{{\tilde W}}$ (pink line),
    $10^{-4}m_{{\tilde W}}$ (blue line), $10^{-3}m_{{\tilde W}}$ (green line),
    $10^{-2}m_{{\tilde W}}$ (light blue line),
    and $T\gg m_{{\tilde W}}$ (red line).  For comparison, we also show
    the annihilation cross section of neutral Wino without the
    Sommerfeld effect (black line).  The vertical axis is the the Wino mass,
    and we take $\Delta m_{{\tilde W}}=160\ {\rm MeV}$.}
\end{figure}

In Fig.\ \ref{fig:cs}, we show the effective annihilation cross
section for several cosmic temperatures, $T= 10^{-5}m_{{\tilde W}},\
10^{-4}m_{{\tilde W}},\ 10^{-3}m_{{\tilde W}}$, and $T\gg m_{{\tilde
    W}}$.  (Here and hereafter, we take $\Delta m_{{\tilde W}}=160\
{\rm MeV}$ in our numerical calculation; even if we vary $\Delta
  m_{{\tilde W}}$ by $5\ {\rm MeV}$ or so, the resultant value of the
  Wino abundance is almost the same. )  For
comparison, we also show the annihilation cross section of neutral
Wino without the Sommerfeld effect in the same figure.  Obviously, the
cross section is significantly enhanced in particular for the Wino
mass larger than $\sim 1\ {\rm TeV}$, and the enhancement factor
becomes larger for lower temperature.  In addition, we can see the
resonance structure at around $m_{{\tilde W}}\simeq 2.4\ {\rm TeV}$,
which occurs due to the existence of zero-energy bound states.  The
precise position of the resonance is determined by the structure of
the electroweak potential, and depends on the mass difference between
charged and neutral Winos, $\Delta m_{\tilde W}$.  For larger $\Delta
m_{\tilde W}$, the resonance peak is shifted to the heavier Wino mass.

As is seen from the Fig.\ \ref{fig:cs}, the annihilation cross section
at the decay temperature can be significantly enhanced by the
Sommerfeld effect.  We stress here again that once coannihilation and
Sommerfeld effects are included, the cross sections show non-trivial
dependence on $T$.  In order to precisely take into account these
effects, we solve the Boltzmann equations numerically.

The relic density of the neutral Wino can be calculated by solving
Eqs.\ \eqref{nwinodot} $-$ \eqref{rhoraddot} with relevant initial
condition.  The initial values of $n_{\tilde{W}}$, $n_X$, and
$\rho_{\rm rad}$ depend on cosmological scenarios and the properties
of $X$.  In the next section, we consider several well-motivated
scenarios and calculate the relic density by numerically solving the
Boltzmann equations.

\section{Relic Abundance of Neutral Wino}
\label{sec:numerical}

\subsection{Case with $X$ domination in the early universe}

First, we consider the case where the particle $X$ once dominates the
universe.  In such a case, at the cosmic time $t\ll \Gamma_X^{-1}$,
the energy density of $X$ is much larger than that of radiation, while
the energy density of radiation (i.e., so-called ``dilute plasma'')
scales as $a^{-3/8}$ (with $a$ being the scale factor).  With such an
initial condition, we solve the Boltzmann equations to calculate the
relic abundance of $\tilde{W}^0$.  In the parameter region of our analysis the maximal
temperature of the dilute plasma is much higher than the Wino mass;
then, the pair annihilation rate is initially much larger than the
expansion rate of the universe.  In such a case, the production and
annihilation terms in the right-hand side of Eq.\ \eqref{nwinodot}
(almost) balance, so we take the initial value of the number density
of the Wino as $n_{\tilde{W}}^{\rm (init)} = \sqrt{n_{\tilde{W}, {\rm
      eq}}^2+N_{\tilde W} \Gamma_X n_X\langle \sigma_{\rm eff} v
  \rangle^{-1}}$.  (However, the resultant relic abundance is
insensitive to the initial value of $n_{\tilde{W}}$ as far as the
initial condition is set at the cosmic time with the background
temperature much higher than $m_{\tilde{W}}$.)  With the initial
condition given above, we calculate the number density of the Wino
after the completion of the decay of $X$, and evaluate the yield
variable of the Wino
\begin{eqnarray}
  Y_{\tilde{W}} \equiv \frac{n_{\tilde{W}}}{s},
\end{eqnarray}
where $s$ is the entropy density.  Using the fact that $Y_{\tilde{W}}$
becomes constant of time at low enough temperature, we calculate the
density parameter as
\begin{eqnarray}
  \Omega_{\tilde{W}} = m_{\tilde{W}}
  \left[ Y_{\tilde{W}} \right]_{t\gg \Gamma_X^{-1}}
  \left( \frac{\rho_{\rm crit}}{s_{\rm now}} \right)^{-1},
\end{eqnarray}
where $\rho_{\rm crit}$ is the critical density of the universe while
$s_{\rm now}$ is the present entropy density, and their ratio is given
by $\rho_{\rm crit}/s_{\rm now} \simeq 3.6 h^2 \times 10^{-9} {\rm
  GeV}$, with $h$ being the Hubble constant in units of $100\ {\rm
  km/sec/Mpc}$.  (In our numerical calculation, we use $h=0.697$
\cite{Hinshaw:2012fq}.)  Comparing $\Omega_{\tilde{W}}$ with the
present density parameter of dark matter, we derive constraints on the
model parameters.  We use the following value of the dark matter
density as the canonical value \cite{Hinshaw:2012fq}:
\begin{eqnarray}
  \Omega_{\rm c} h^2 = 0.1146.
\end{eqnarray}

In the case where $X$ once dominates the universe, the relic density
of the Wino depends on the following parameters: $\Gamma_X$ (decay
rate of $X$), $m_X$ (mass of $X$), $m_{\tilde{W}}$ (Wino mass), and
$N_{\tilde{W}}$ (averaged number of the Winos produced by the decay of
one $X$).  In particular, the relic density depends on the decay rate
$\Gamma_X$.  To discuss the dependence on $\Gamma_X$, it is convenient
to define the ``decay temperature'' as
\begin{eqnarray}
  T_X \equiv
  \left(
    \frac{10}{g_*(T_X) \pi^2} M_{\rm Pl}^2 \Gamma_{X}^2
  \right)^{1/4},
  \label{T_R}
\end{eqnarray}
where $M_{\rm Pl}\simeq 2.4\times 10^{18}\ {\rm GeV}$ is the reduced
Planck scale.  Notice that $T_X$ corresponds to the cosmic temperature
at the time of $X$ decay.

\begin{figure}[t]
  \centerline{\epsfxsize=0.75\textwidth\epsfbox{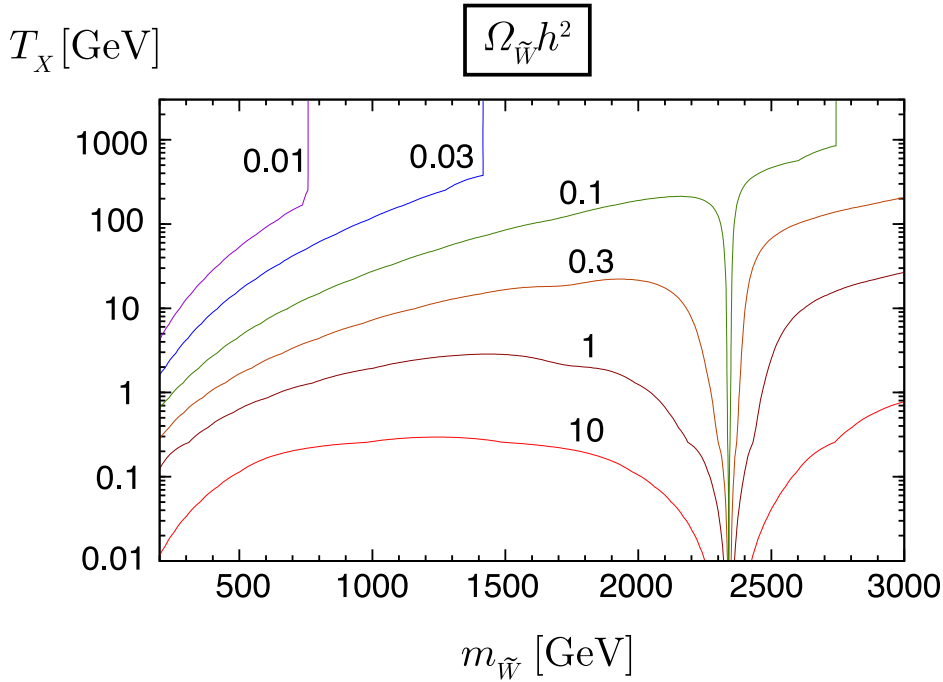}}
  \caption{Contours of constant $\Omega_{\tilde{W}}h^2$ for
    $N_{\tilde{W}}=1$ and $m_X=5m_{\tilde{W}}$ for the case where $X$
    once dominates the universe.  Numbers in the figure are
    $\Omega_{\tilde{W}}h^2$.}
  \label{fig:N1moju}
\end{figure}

In Fig.\ \ref{fig:N1moju}, we show the contours of constant
$\Omega_{\tilde{W}}h^2$ for the case with $N_{\tilde{W}}=1$. 
  (In this subsection, we take $m_X/m_{\tilde{W}}=5$; even if we vary
  this ratio, the resultant value of $\Omega_{\tilde{W}}h^2$ is almost
  unchanged if the annihilation of the Winos becomes effective.)  As
one can see, $\Omega_{\tilde{W}}h^2$ is insensitive to $T_X$ when the
decay temperature is high, while $\Omega_{\tilde{W}}h^2$ increases as
$T_X$ becomes lower when the decay temperature is relatively low.
Such behaviors can be easily understood.  With high enough decay
temperature, the Winos are still in chemical equilibrium after the
completion of the $X$ decay, and hence the relic abundance of the Wino
is given by the thermal relic density.  On the contrary, with low
decay temperature, the Winos cannot be in chemical equilibrium.  Even
in such a case, the pair annihilation of the Wino proceeds as far as
the annihilation rate is larger than the expansion rate of the
universe.  In such a case, the yield value of the Wino is approximated
by
\begin{eqnarray}
  Y_{\tilde{W}} \sim
  {\rm min} \left[
    \frac{3\Gamma_X}{\langle \sigma_{\rm eff}  v \rangle s(T_X)}, 
    \frac{N_{\tilde W} n_X (T_X)}{s(T_X)}
  \right],
  \label{Ywino(approx)}
\end{eqnarray}
where $n_X (T_X)$ represents the number density of $X$ just before the
decay.  
 Notice that the first (second) term in the
  right-hand side of Eq.\ \eqref{Ywino(approx)} is relevant for the
  case where the effect of the annihilation
  of the Wino is effective (ineffective).
The first term, which is approximately
proportional to $T_X^{-1}$, is smaller than the second one, resulting
in the enhancement of $\Omega_{\tilde{W}}h^2$ with lower decay
temperature.

In order to realize the Wino dark matter scenario (i.e.,
$\Omega_{\tilde{{W}}}=\Omega_{\rm c}$), relatively high value of $T_X$
is needed.  For $m_{\tilde{W}}=300\ {\rm GeV}$ ($500\ {\rm GeV}$, $1\
{\rm TeV}$, $2\ {\rm TeV}$), $T_X= 1.3\ {\rm GeV}$ ($4.0\ {\rm GeV}$,
$23\ {\rm GeV}$, $150\ {\rm GeV}$) is necessary.  If $T_X$ is lower,
the universe is overclosed.  This has an important implication when
the cosmological moduli fields play the role of $X$, as we discuss
below.  In addition, we can see a significant suppression of
$\Omega_{\tilde{W}}h^2$ when $m_{\tilde{W}}\simeq 2.4\ {\rm TeV}$.
Such a suppression is due to the significant enhancement of the
annihilation cross section by the Sommerfeld effect.

\begin{figure}[t]
  \centerline{\epsfxsize=0.75\textwidth\epsfbox{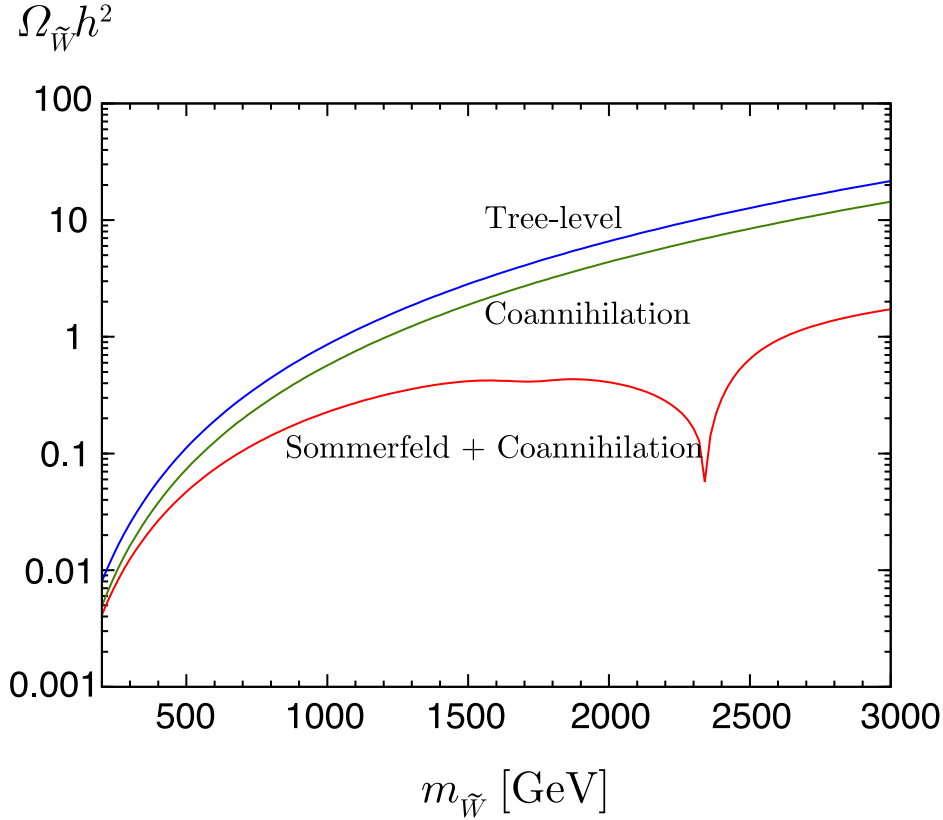}}
  \caption{${\Omega}_{\tilde{W}}h^2$ as a function of $m_{\tilde{W}}$
    for $N_{\tilde{W}}=1$ and $T_X=10\ {\rm GeV}$ with Sommerfeld and
    coannihilation effects (red line), with coannihilation effect only
    (green line) and without neither Sommerfeld nor coannihilation
    effects (blue line).
}
\label{fig:hikaku_eff}
\end{figure}

In order to see the importance of Sommerfeld enhancement, we also
calculate $\Omega_{\tilde{W}}h^2$ neglecting the Sommerfeld
enhancement.  In addition, for comparison, we also show the result
without the effect of coannihilation, for which we use
$\langle\sigma_{\rm eff}v\rangle=
\langle\sigma_{\tilde{W}^0\tilde{W}^0\rightarrow W^+W^-} v\rangle$
(without Sommerfeld effect).  The results are shown in Fig.\
\ref{fig:hikaku_eff} for $N_{\tilde{W}}=1$ and $T_X=10\ {\rm GeV}$.
The Sommerfeld effect significantly changes ${\Omega}_{\tilde{W}}h^2$
when $m_{\tilde{W}}\gtrsim 1\ {\rm TeV}$.

\begin{figure}[t]
  \centerline{\epsfxsize=0.75\textwidth\epsfbox{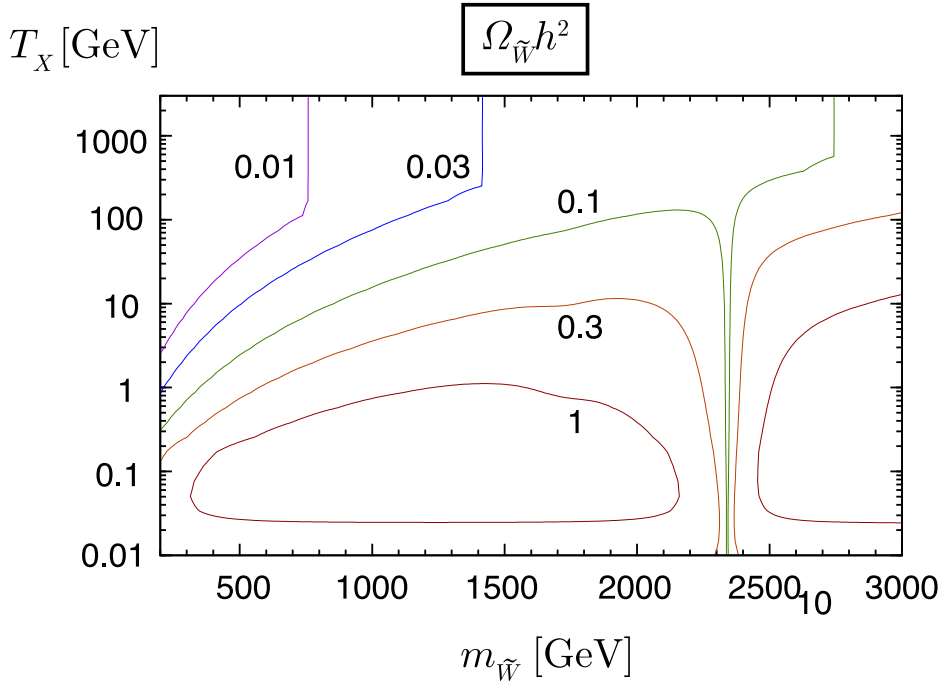}}
  \caption{Same as Fig.\ \ref{fig:N1moju}, except for
    $N_{\tilde{W}}=10^{-6}$.}
  \label{fig:N04moju}
\end{figure}

\begin{figure}[t]
  \centerline{\epsfxsize=0.75\textwidth\epsfbox{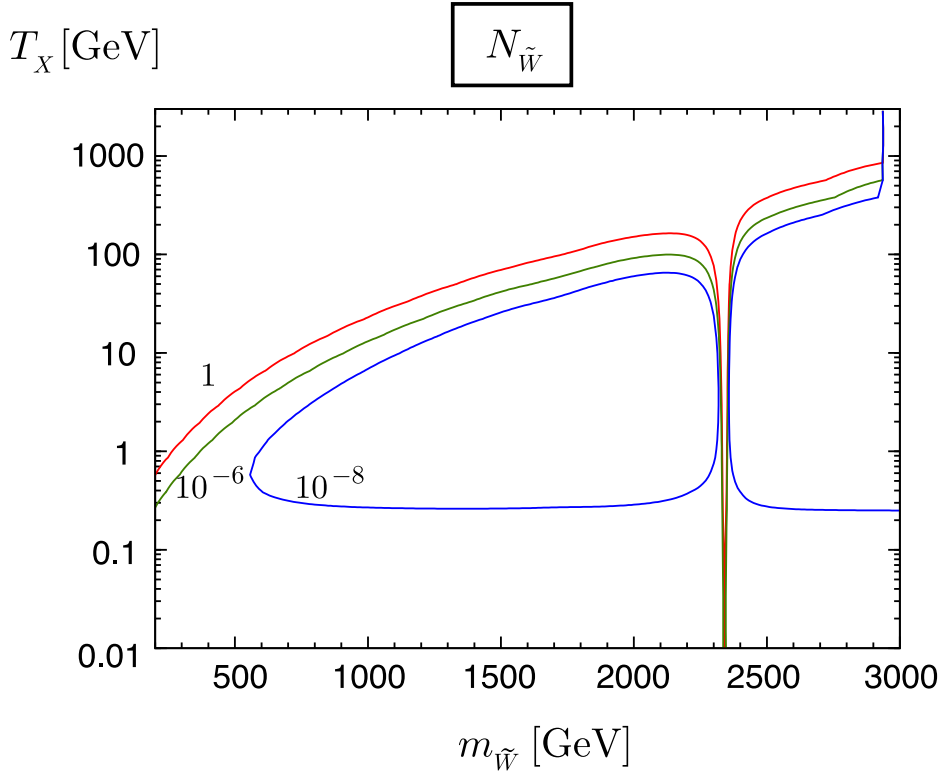}}
  \caption{Contours of ${\Omega}_{\tilde{W}}=\Omega_c$ for
    $N_{\tilde{W}}=1$ (blue line), $N_{\tilde{W}}=10^{-6}$ (green
    line) and $N_{\tilde{W}}=10^{-8}$ (red line).}
  \label{fig:mhi}
\end{figure}

In the case where $X$ dominates the universe, the $N_{\tilde{W}}$
dependence of $\Omega_{\tilde{W}}h^2$ is very weak unless
$N_{\tilde{W}}$ is significantly suppressed.  In Fig.\
\ref{fig:N04moju}, we plot contours of constant
${\Omega}_{\tilde{W}}h^2$ with $N_{\tilde{W}}=10^{-6}$.  Even with
such a small value of $N_{\tilde{W}}$, the contour of
$\Omega_{\tilde{W}}=\Omega_c$ does not change much compared to that
given in Fig.\ \ref{fig:N1moju}.  Thus, unless $N_{\tilde{W}}\ll
10^{-6}$, $T_X$ should be higher than $0.6\ {\rm GeV}$ ($2\ {\rm
  GeV}$, $10\ {\rm GeV}$, and $90\ {\rm GeV}$) for $m_{\tilde{W}}=300\
{\rm GeV}$ ($m_{\tilde{W}}=500\ {\rm GeV}$, $m_{\tilde{W}}=1\ {\rm
  TeV}$, and $m_{\tilde{W}}=2\ {\rm TeV}$).  In Fig.\ \ref{fig:mhi},
we show the contours of the upper bound on $N_{\tilde{W}}$ on the
$m_{\tilde{W}}$ vs.\ $T_X$ plane, requiring
$\Omega_{\tilde{W}}<\Omega_{\rm c}$.  As a result, we can see that
$N_{\tilde{W}}\lesssim O(10^{-8})$ is required if $T_X$ is so small
that the pair annihilation of the Winos is ineffective.

One of the important applications of our study is to the case with
weakly interacting late decaying scalar fields, like scalar fields
responsible for the SUSY breaking (i.e., so-called the Polonyi field)
and the moduli fields in string theory.  (Hereafter, we call them
``moduli fields.'')  Moduli fields are expected to have interactions
suppressed by the Planck scale.  A modulus field may have a large
initial amplitude.  Because of the weakness of the interaction, the
lifetime of the modulus may be so long that its decay occurs at a late
stage of the evolution of the universe.  It is well known that the
late-time decay of the modulus field is dangerous because it destroys
the light elements produced by the BBN reactions, resulting in the
spoil of the success of the BBN scenario.  Such a problem can be
avoided if the decay rate of the modulus is somehow enhanced so that
the lifetime of the modulus becomes shorter than $\sim 1\ {\rm sec}$.
Even so, the decay of the modulus field produces significant amount of
superparticles, which may result in the overproduction of the LSP.
With the Wino LSP, this problem may be avoided.

The decay rate of a modulus field depends on how it interacts with the
MSSM fields.  For example, it may interact as
\begin{eqnarray}
  {\cal L}_{\rm int} = 
  \frac{\lambda_{\rm G}}{M_{\rm Pl}}
  \int d^2 \theta
  \hat{X} \hat{\cal W}^\alpha \hat{\cal W}_\alpha + {\rm h.c.},
  \label{XWW}
\end{eqnarray}
where the ``hat'' stands for superfields, $\hat{\cal W}$ is the gauge
field strength superfield, and $\lambda_{\rm G}$ is a coupling
constant.  Then, the scalar component $X$ decays into a pair of vector
boson $V$ with the following decay rate:
\begin{eqnarray}
  \Gamma_{X\rightarrow VV} &=& 
  \frac{N_{\rm f}\lambda_{\rm G}^2}{4\pi}
  \frac{m_X^3}{M_{\rm Pl}^2},
\end{eqnarray}
where $N_{\rm f}$ is the number of the possible final states; for
example, $N_{\rm f}=N^2-1$ for an SU($N$) gauge group.  (Here, we
assume that the vector boson is much lighter than $X$.)  In addition,
with the interaction given in Eq.\ \eqref{XWW}, $X$ may decay into a
pair of gauginos $\lambda$.  The decay rate is model-dependent, and is
typically $\Gamma_{X\rightarrow \lambda \lambda}\sim
\Gamma_{X\rightarrow VV}$ \cite{Endo:2006zj, Nakamura:2006uc}.
Approximating $\Gamma_X\sim\Gamma_{X\rightarrow VV}$, the decay
temperature is estimated to be
\begin{eqnarray}
  T_X \sim 0.01\ {\rm GeV} \times
  \lambda_{\rm G}
  \left( \frac{m_X}{100\ {\rm TeV}} \right)^{3/2},
\end{eqnarray}
where we have used $N_{\rm f}=12$, which is the number of the gauge
bosons in the standard model.  For $m_{\tilde{W}}=300\ {\rm GeV}$ ($1\
{\rm TeV}$), the $\lambda_{\rm G}$-parameter is required to be larger
than $430$, $130$, and $42$ ($7600$, $2300$, and $74$) for $m_X=10\
{\rm TeV}$, $100\ {\rm TeV}$, and $1000\ {\rm TeV}$, respectively,
where we have assumed $N_{\tilde{W}}\sim 1$.  It seems that if such a
modulus with ${\lambda}_{G}\sim O(1)$ once dominates the universe it
tends to cause overproduction of the Winos.  In order for the Winos
not to overclose the universe, large $m_X$ or enhanced $\lambda_{\rm
  G}$ is needed.

\subsection{Case with gravitino}

Next, we consider the case where gravitino plays the role of $X$.
Even though the gravitino may not dominate the universe, a significant
amount of $\tilde{W}$ may be produced by its decay.

The gravitino is the superpartner of graviton, and it couples to the
supercurrent so it interacts with all the supermultiplets.  If
unstable, the gravitino decays into an ordinary (visible-sector)
particle and its superpartner.  Thus, decay of the gravitino results
in the production of the LSP because all the produced superparticles
cascade down to the LSP.  The interactions of the gravitino are
suppressed by inverse powers of the Planck scale so that the decay
rate of the gravitino is extremely small.  It is well known that the
late-time decay of the gravitino may spoil the success of the BBN if
the lifetimes of the gravitino is longer than $\sim 1\ {\rm sec}$
\cite{Kawasaki:2004yh, Kawasaki:2004qu, Kawasaki:2008qe}.  In order to
avoid such a problem, we concentrate on the case where the lifetime is
shorter than $\sim 1\ {\rm sec}$, which is realized if the gravitino
mass is larger than $\sim O(10\ {\rm TeV})$.  Assuming that the
gaugino masses are at the TeV scale and are much smaller than the
gravitino mass, the lifetime of the gravitino is estimated as
\begin{eqnarray}
  \tau_{3/2} = 0.4\ {\rm sec} \times N_{\rm G}^{-1}
  \left(\frac{m_{3/2}}{100\ {\rm TeV}}\right)^{-3},
\end{eqnarray}
where $m_{3/2}$ is gravitino mass.  In addition, $N_{\rm G}$ is the
number of gauge multiplets to which the gravitino decays.  (In the
following numerical study, we take the MSSM value of $N_{\rm G}=12$.)
We also assume that the superparticles other than gauginos are as
heavy as the gravitino so that the decay modes into those
superparticles are negligible.

Even though the gravitino is very weakly interacting, gravitinos are
produced by scattering processes of particles in thermal bath.  The
abundance of the gravitino (before its decay) is approximately
proportional to the reheating temperature after inflation; for the
case where the gravitino mass is significantly larger than gaugino
masses, the yield variable of the gravitino, which is defined as
$Y_{3/2}=n_{3/2}/s$ with $n_{3/2}$ being the number density of the
gravitino, is given by \cite{Kawasaki:2008qe}\footnote
{If there exists condensation of a scalar field (like inflaton, moduli
  fields, and so on), the gravitino may be produced by the decay
  process \cite{Endo:2006zj, Nakamura:2006uc, Kawasaki:2006gs}.  We
  assume that the abundance of the gravitino from such a decay
  process, which is highly model-dependent, is negligible.}
\begin{eqnarray}
  \left[ Y_{3/2} \right]_{t\ll \tau_{3/2}}
  \simeq
  2.3 \times 10^{-14} \times T_{\rm R}^{(8)}
  \left[ 1 + 0.015 \ln T_{\rm R}^{(8)} - 0.0009 \ln^2 T_{\rm R}^{(8)}
    \right],
  \label{Ygrav}
\end{eqnarray}
where $T_{\rm R}^{(8)}\equiv T_{\rm R}/10^8\ {\rm GeV}$, with $T_{\rm
  R}$ being the reheating temperature after inflation,\footnote
{Here, we assume that there is no significant entropy production after
  inflation.}
which is related to the decay rate of the inflaton $\Gamma_{\rm inf}$
as
\begin{eqnarray}
  T_{\rm R} \equiv
  \left(
    \frac{10}{g_*(T_{\rm R}) \pi^2} M_{\rm Pl}^2 \Gamma_{\rm inf}^2
  \right)^{1/4}.
\end{eqnarray}

\begin{figure}[t]
  \centerline{\epsfxsize=0.75\textwidth\epsfbox{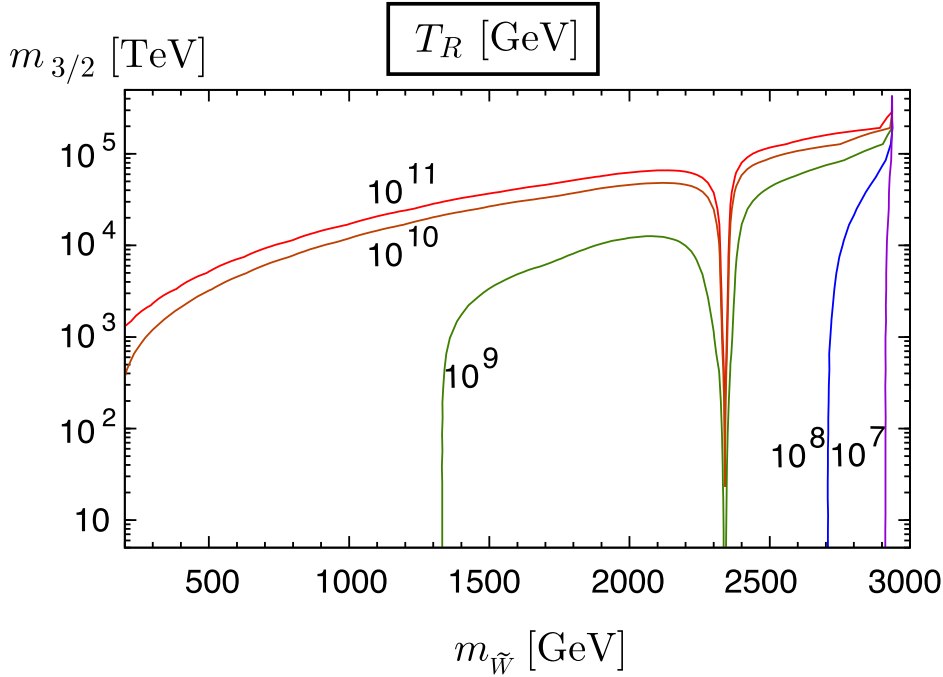}}
  \caption{Contours of $\Omega_{\tilde{W}}=\Omega_c$ for several
    values of the reheating temperature.  Here we take $N_{\rm G}=12$.
    Numbers in the figure are $T_{\rm R}$ in units of GeV.}
  \label{fig:grvn}
\end{figure}

We calculate the density parameter of the Wino by numerically solving
the Boltzmann equations given in Section \ref{sec:formulas}.  Because
the primordial abundance of the gravitino is approximately
proportional to the reheating temperature, $\Omega_{\tilde{W}}$ also
increases with higher reheating temperature.

\begin{figure}[t]
  \centerline{\epsfxsize=0.75\textwidth\epsfbox{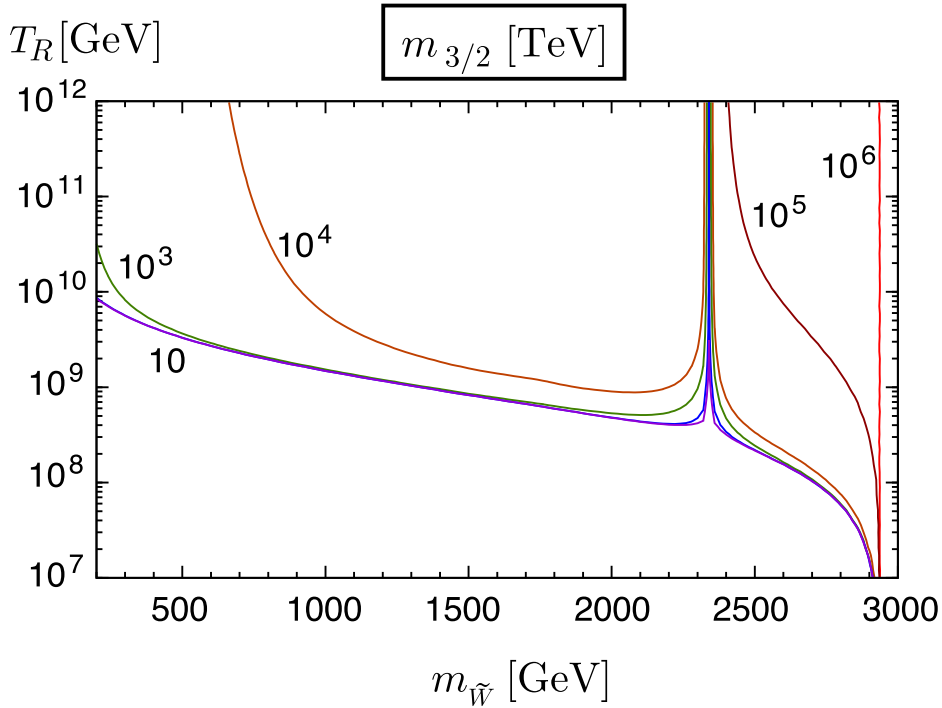}}
  \caption{Upper bound on the reheating temperature as a function of
    the Wino mass for several values of the gravitino mass.  Numbers
    in the figure are the gravitino mass in units of TeV.  (The bound
    for $m_{3/2}=10^2\ {\rm TeV}$ (blue line) almost overlaps with
    that of $m_{3/2}=10\ {\rm TeV}$.)}
  \label{fig:grvnTR}
\end{figure}

In Fig.\ \ref{fig:grvn}, we show the contours of
$\Omega_{\tilde{W}}=\Omega_c$ for several values of $T_{\rm R}$ on
$m_{\tilde{W}}$ vs.\ $m_{3/2}$ plane.\footnote
{In the anomaly-mediation scenario with $m_{\tilde{W}}\sim O(100\ {\rm
    GeV}-1\ {\rm TeV})$, the gravitino mass is expected to be of
  $O(100\ {\rm TeV})$.  Here, however, we vary $m_{3/2}$ up to
  $O(10^5\ {\rm TeV})$ to provide information about the case with
  extremely heavy gravitino.}
The contour of $\Omega_{\tilde{W}}=\Omega_{\rm c}$ is almost
independent of the reheating temperature if $T_{\rm R}$ is high
enough.  This is due to the fact that, with high enough reheating
temperature and large enough gravitino mass, the Wino abundance at the
time of the gravitino decay is so large that the pair annihilation of
Wino becomes effective.  Then, as discussed in the previous
subsection, $\Omega_{\tilde{W}}$ is determined by the decay
temperature which is determined by $m_{3/2}$ and becomes insensitive
to the primordial abundance of the gravitino.  On the contrary, if the
gravitino mass is smaller than $\sim 10^3\ {\rm TeV}$, the contour of
$\Omega_{\tilde{W}}=\Omega_{\rm c}$ becomes insensitive to the
gravitino mass.  This is because, if the gravitino mass is small, the
decay temperature of the gravitino becomes so low that the
annihilation of the Wino is inefficient.  Then, almost all Winos
produced by the gravitino decay survive until today, and the relic
density of the Wino is approximately given by the sum of thermal relic
density and the non-thermal one from the gravitino decay:
$Y_{\tilde{W}}\simeq Y_{\tilde{W}}^{\rm (th)}+[Y_{3/2}]_{t\ll
  \tau_{3/2}}$, with $Y_{\tilde{W}}^{\rm (th)}$ being the thermal
abundance of the Wino.  The thermal relic density becomes larger than
the present dark matter density if $m_{\tilde{W}}\gtrsim 2.9\ {\rm
  TeV}$, so the Wino mass larger than $\sim 2.9\ {\rm TeV}$ is
forbidden irrespective of $T_{\rm R}$.  On the contrary, with smaller
Wino mass, we obtain upper bound on the reheating temperature by
requiring $\Omega_{\tilde{W}}<\Omega_c$, using the fact that the
primordial number density of the gravitino is approximately
proportional to $T_{\rm R}$.  Such a bound is shown in Fig.\
\ref{fig:grvnTR} for several values of the gravitino mass.

One implication of our result is on the leptogenesis scenario
\cite{Fukugita:1986hr}.  The leptogenesis scenario requires the
reheating temperature to be higher than $\sim 10^9\ {\rm GeV}$ in
order to generate large enough amount of the baryon asymmetry of the
universe \cite{Buchmuller:2004nz, Giudice:2003jh}.  We have seen that
such a high reheating temperature can be realized in the Wino LSP
case.

\subsection{Case with SUSY Peccei-Quinn model}

The Peccei-Quinn (PQ) mechanism \cite{Peccei:1977hh, Peccei:1977ur} is
an attractive solution to the strong CP problem. If we embed the PQ
mechanism into supersymmetric model, there exists the fermionic
superpartner of the axion called axino. Axinos are copiously produced
in the thermal bath of early universe and decay into MSSM particles at
a late stage of the cosmic expansion.  In this subsection, we consider
the case where the axino $\tilde{a}$ plays the role of $X$, which is
embedded in the axion multiplet $\hat{\cal A}$ as\footnote
{The real scalar component of $\hat{\cal A}$ (i.e., saxion) may also
  cause cosmological difficulties.  Cosmology with saxion depends on
  the model of PQ sector as well as on thermal history.  See
  \cite{Kim:2008yu, Kawasaki:2010gv, Kawasaki:2011ym, Moroi:2012vu,
    Mukaida:2012qn} for more detailed discussion.}
\begin{eqnarray}
  \hat{\cal A} = \frac{1}{\sqrt{2}} (\sigma + ia) 
  + \sqrt{2} \theta \tilde{a}
  + (\mbox{$F$-term}).
\end{eqnarray}
(In this subsection, the field $X$ is denoted as $\tilde{a}$.)

To make our discussion concrete, we assume that the PQ charges of all
the MSSM fields are zero, and that the PQ fermions are embedded into
full multiplets of $SU(5)$ grand unified gauge group.  In such a
model, denoting the PQ scale as $f_a$, the axion multiplet $\hat{\cal
  A}$ has the following interaction term:
\begin{eqnarray}
  {\cal L}_{\rm int} = \frac{1}{4\sqrt{2}\pi f_a}
  \int d^2 \theta
  \left[
    \alpha_3 \hat{\cal A} \hat{\cal G}^\alpha \hat{\cal G}_\alpha 
    + \alpha_2 \hat{\cal A} \hat{\cal W}^\alpha \hat{\cal W}_\alpha 
    + \frac{5}{3}
    \alpha_1 \hat{\cal A} \hat{\cal B}^\alpha \hat{\cal B}_\alpha
  \right]
  + {\rm h.c.},
  \label{AWW}
\end{eqnarray}
where $\hat{\cal G}$, $\hat{\cal W}$, and $\hat{\cal B}$ are
field strength superfields for $SU(3)_C$, $SU(2)_L$, and $U(1)_Y$,
respectively, and $\alpha_a\equiv g_a^2/4\pi$ with $g_a$ ($a=1-3$)
being gauge coupling constants.  (The summation over the gauge indices
are implicit for $SU(3)_C$ and $SU(2)_L$.)

As in the case of the gravitino, the primordial abundance of the axino
depends on the reheating temperature after inflation
\cite{Covi:2001nw}.  We adopt the yield variable of axino (i.e.,
$Y_{\tilde{a}}=n_{\tilde{a}}/s$, with $n_{\tilde{a}}$ being the number
density of axino) evaluated in \cite{Brandenburg:2004du}:
\begin{eqnarray}
  \left[ Y_{\tilde{a}} \right]_{t\ll \Gamma_{\tilde{a}}^{-1}}
  \simeq
  \mbox{min} \left[
    Y_{\tilde{a}}^{\rm (eq)},
    0.20 \times \alpha_3^3 
    \ln \left( \frac{0.0977}{\alpha_3} \right)
    \left( \frac{T_{\rm R}}{10^{7}\ {\rm GeV}} \right)
    \left( \frac{f_a}{10^{11}\ {\rm GeV}} \right)^{-2}
  \right],
  \label{Y(axino)}
\end{eqnarray}
where $Y_{\tilde{a}}^{\rm (eq)}\simeq 1.8\times 10^{-3}$ is the
thermal abundance of axino.  If the axino is stable, the axinos
produced in the early universe survive until today.  Then, in order
not to overclose the universe with the mass density of axino, severe
upper bound on the reheating temperature is obtained
\cite{Covi:2001nw, Brandenburg:2004du, Bae:2011jb, Choi:2011yf}.  Even
if all the LSPs produced by the axino decay remain until today, which
is the case if the annihilation cross section of the LSP is small, a
stringent upper bound on $T_{\rm R}$ still exists.  With a large
annihilation cross section of dark matter, this problem may be avoided
\cite{Choi:2008zq}.

One of the important parameters to calculate the relic Wino abundance
produced by the axino decay is the decay rate of axino.  In the
present setup, the axino decays dominantly into gauge boson and
gaugino pair; with the interaction terms given in Eq.\ \eqref{AWW},
the decay rate is given by
\begin{eqnarray}
  \Gamma_{\tilde{a}} = 
  \Gamma_{\tilde{a}\rightarrow g\tilde{g}} + 
  \Gamma_{\tilde{a}\rightarrow W^\pm\tilde{W}^\mp} + 
  \Gamma_{\tilde{a}\rightarrow Z\tilde{W}^0} + 
  \Gamma_{\tilde{a}\rightarrow \gamma\tilde{W}^0} + 
  \Gamma_{\tilde{a}\rightarrow Z\tilde{B}} + 
  \Gamma_{\tilde{a}\rightarrow \gamma\tilde{B}},
\end{eqnarray}
where, if kinematically allowed \cite{Baer:2011hx},
\begin{eqnarray}
  \Gamma_{\tilde{a}\rightarrow g\tilde{g}} &=&
  \frac{8{\alpha_3}^2}{128{\pi}^3 }\frac{{m_{\tilde{a}}}^3}{{f_a}^2}
  \left( 1 - {y_{{\tilde{g}}}}^2 \right)^3,
  \label{axino2gluino}
  \\
  \Gamma_{\tilde{a}\rightarrow W^\pm\tilde{W}^\mp}  &=&
  \frac{2{\alpha}_2^2}{128{\pi}^3 }\frac{{m_{\tilde{a}}}^3}{{f_a}^2}
  K(y_{\tilde{W}},y_W),
  \\
  \Gamma_{\tilde{a}\rightarrow Z\tilde{W}^0}  &=& 
  \frac{{\alpha}_2^2{\cos}^2{\theta}_{W}}{128 {\pi}^3}
  \frac{{m_{\tilde{a}}}^3}{{f_a}^2}
  K(y_{\tilde{W}},y_{Z}),
  \\
  \Gamma_{\tilde{a}\rightarrow \gamma\tilde{W}^0}  &=& 
  \frac{{\alpha}_2^2{\sin}^2{\theta}_{W}}{128 {\pi}^3 }
  \frac{{m_{\tilde{a}}}^3}{{f_a}^2}
  \left( 1 - {y_{\tilde{W}}}^2 \right)^3 ,
  \\
  \Gamma_{\tilde{a}\rightarrow Z\tilde{B}}  &=&  
  \left( \frac{5}{3} \right)^2
  \frac{{\alpha}_1^2{\sin}^2{\theta}_{W}}{128 {\pi}^3 }
  \frac{{m_{\tilde{a}}}^3}{{f_a}^2}K(y_{\tilde{B}},y_Z),
  \\
  \Gamma_{\tilde{a}\rightarrow \gamma\tilde{B}} &=& 
  \left( \frac{5}{3} \right)^2
  \frac{{\alpha}_1^2 {\cos}^2{\theta}_{W}}{128
  {\pi}^3}
  \frac{{m_{\tilde{a}}}^3}{{f_a}^2} \left( 1 - {y_{\tilde{B}}}^2 \right)^3 ,
  \label{axino2gammabino}
\end{eqnarray}
where ${\theta}_W$ is the Weinberg angle, $y_I\equiv
m_I/m_{\tilde{a}}$ denotes the mass of the particle $I$ normalized by the
axino mass $m_{\tilde{a}}$, and
\begin{eqnarray}
 K(y_1,y_2)\! =\! \sqrt{1+y_1^4+y_2^4-2y_1^2-2y_2^2-2y_1^2 y_2^2}
  \left[ (1-y_1^2)^2\! +3y_1y_2^2-\frac{y_2^2}{2}(1+y_1^2+y_2^2) \right].
\end{eqnarray}

\begin{figure}
  \centerline{\epsfxsize=0.75\textwidth\epsfbox{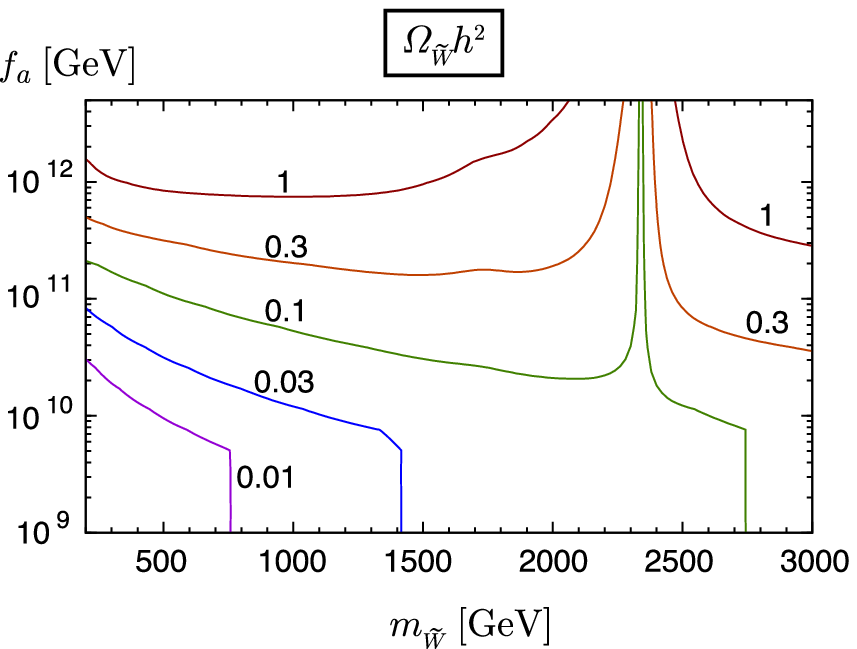}}
  \caption{Contour of $ \Omega_{ \tilde{W}} h^2$ of the present
    universe for mass ratio
    $m_{\tilde{W}}:m_{\tilde{B}}:m_{\tilde{g}}:m_{\tilde{a}}=1:3:7:10$.
    Here, we take $\left[ Y_{\tilde{a}} \right]_{t\ll
      \Gamma_{\tilde{a}}^{-1}}=Y_{\tilde{a}}^{\rm (eq)}$.  Numbers in
    the figure are the values of $\Omega_{\tilde{W}}h^2$.}
  \label{Fig:an7_10}
%
  \centerline{\epsfxsize=0.75\textwidth\epsfbox{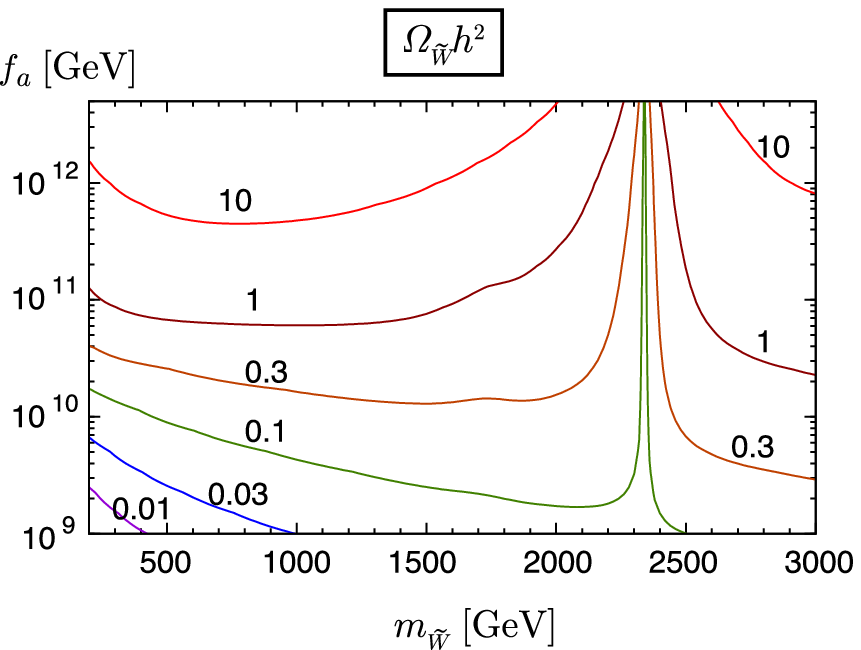}}
  \caption{Same as Fig.\ \ref{Fig:an7_10}, except for
    $m_{\tilde{W}}:m_{\tilde{B}}:m_{\tilde{g}}:m_{\tilde{a}}=1:3:7:5$.}
  \label{Fig:an7_5}
\end{figure}

The decay temperature, and hence the resultant Wino abundance, depend
on the mass spectrum of superparticles, in particular, that of
gauginos.  In the case where the gluino is lighter than the axino, the
axino dominantly decays into gluon and gluino pair.  In such a case,
the decay rate of the axino becomes relatively large.  In the opposite
case, the dominant decay modes of the axino are $\tilde{a}\rightarrow
W^\pm\tilde{W}^\mp$, $\tilde{a}\rightarrow Z\tilde{W}^0$, and
$\tilde{a}\rightarrow \gamma\tilde{W}^0$, and the decay rate of the
axino is suppressed.  So, in the latter case, $\Omega_{\tilde{W}}$
becomes larger compared to the former case.

In Fig.\ \ref{Fig:an7_10}, we show the contours of constant
$\Omega_{\tilde{W}}h^2$ for the case with high enough reheating
temperature ($ \left[ Y_{\tilde{a}} \right]_{t\ll
  \Gamma_{\tilde{a}}^{-1}}=Y_{\tilde{a}}^{\rm (eq)}$), taking
$m_{\tilde{W}}:m_{\tilde{B}}:m_{\tilde{g}}:m_{\tilde{a}}=1:3:7:10$.
(The gaugino masses are assumed to obey the anomaly-mediation relation
\cite{Giudice:1998xp, Randall:1998uk}.)  With such a mass spectrum,
the axino dominantly decays into gluon and gluino pair.  Then,
$\Omega_{\tilde{W}}=\Omega_c$ is realized with $f_a=1.9\times10^{11}\
{\rm GeV}$ ($1.3\times10^{11} {\rm GeV}$, $6.3\times10^{10} {\rm
  GeV}$, $2.7\times 10^{10} {\rm GeV}$) for $m_{\tilde{W}}=300\ {\rm
  GeV}$ ($m_{\tilde{W}}=500\ {\rm GeV}$, $m_{\tilde{W}}=1\ {\rm TeV}$,
and $m_{\tilde{W}}=2\ {\rm TeV}$).

If we consider the case where the decay mode into the gluon and gluino
pair is kinematically blocked, the decay rate is suppressed. In such a
case, the PQ scale which realizes the Wino dark matter becomes
smaller.  In Fig.\ \ref{Fig:an7_5}, we show the result taking
$m_{\tilde{W}}:m_{\tilde{B}}:m_{\tilde{g}}:m_{\tilde{a}}=1:3:7:5$.
Then, the value of $f_a$ giving rise to $\Omega_{\tilde{W}}=\Omega_c$
is given by $1.5\times 10^{10} {\rm GeV}$ ($1.0\times 10^{10} {\rm
  GeV}$, $5.1\times 10^9 {\rm GeV}$, $2.2\times 10^9 {\rm GeV}$) for
$m_{\tilde{W}}=300\ {\rm GeV}$ ($m_{\tilde{W}}=500\ {\rm GeV}$,
$m_{\tilde{W}}=1\ {\rm TeV}$, and $m_{\tilde{W}}=2\ {\rm TeV}$).

\begin{figure}[t]
  \centerline{\epsfxsize=0.75\textwidth\epsfbox{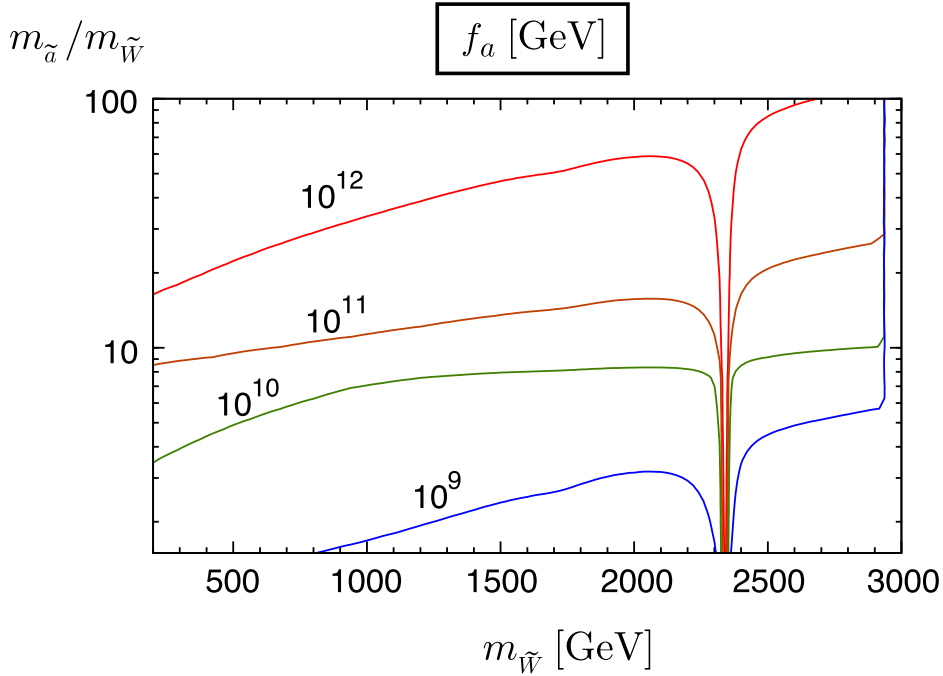}}
  \caption{Contours of constant $f_a$ which gives
    $\Omega_{\tilde{W}}=\Omega_c$ for $\left[ Y_{\tilde{a}}
    \right]_{t\ll \Gamma_{\tilde{a}}^{-1}}=Y_{\tilde{a}}^{\rm (eq)}$.
    The horizontal axis is the Wino mass while the vertical one is the
    ratio $m_{\tilde{a}}/m_{\tilde{W}}$.  The gaugino masses are
    assumed to obey the anomaly-mediation relation
    $m_{\tilde{W}}:m_{\tilde{B}}:m_{\tilde{g}}=1:3:7$.}
  \label{Fig:famax}
\end{figure}

We have also studied how the required value of $f_a$ to realize the
Wino dark matter depends on the axino mass.  In Fig.\ \ref{Fig:famax},
we show contours of constant $f_a$ which gives
$\Omega_{\tilde{W}}=\Omega_c$ for $\left[ Y_{\tilde{a}} \right]_{t\ll
  \Gamma_{\tilde{a}}^{-1}}=Y_{\tilde{a}}^{\rm (eq)}$ on
$m_{\tilde{W}}$ vs.\ $m_{\tilde{a}}/m_{\tilde{W}}$ plane.

So far, we have adopted the thermal abundance of axino.  Even if the
primordial abundance of axino is smaller, the resultant Wino density
does not change as far as Winos produced by decay of axino are
much enough to pair-annihilate.  We also performed the calculation
with lower reheating temperature, and checked that the results are
more or less unchanged.

\section{Conclusions and Discussion}
\label{sec:conclusions}

In this paper, we have studied the possibility of the Wino cold dark
matter in supersymmetric models, paying particular attentions to the
scenario in which the decay of a long-lived particle $X$ produces
significant amount of the Wino LSP.  We have numerically calculated
the relic abundance of the neutral Wino, carefully taking account of
the effects of coannihilation and Sommerfeld effects.  We have seen
that the Sommerfeld effect drastically enhances the pair annihilation
cross section of the Wino if $m_{\tilde{W}}\gtrsim 1\ {\rm TeV}$.  We
have studied the cases where moduli fields, gravitino, or axino plays
the role of $X$, and derived the constraints on the model parameters
in each case.

So far, we have not discussed cosmological and astrophysical
constraints on the Wino dark matter.  Because the neutral Wino has
large annihilation cross section, the Wino dark matter scenario
conflicts with astrophysical and cosmological constraints if the Wino
mass is too small.  One of the constraints is from the negative
observation of high energy $\gamma$-ray from dwarf galaxies, from
which the mass regions $m_{\tilde{W}}\lesssim 400\ {\rm GeV}$ are
disfavored \cite{Ackermann:2011wa, Abazajian:2011ak} although
astrophysical uncertainties exist.  Another constraint is from
BBN. The relic Wino may pair annihilated during and after the BBN
epoch, which causes photo- and hadro-dissociation processes.  In order
not to spoil the success of the BBN scenario, the Wino mass is
required to be larger than $\sim 200\ {\rm GeV}$ \cite{Hisano:2008ti}.
Thus, these astrophysical and cosmological constraints do not exclude
most of the parameter regions we have studied.

\vspace{1em}
\noindent {\it Acknowledgements}: This work is supported in part by
Grant-in-Aid for Scientific research from the Ministry of Education,
Science, Sports, and Culture (MEXT), Japan, No.\ 22244021, No.\
22540263, and No.\ 23104008.

\appendix
\section{Sommerfeld Enhancement Factors}
\setcounter{equation}{0}

\begin{table}
\begin{center}
\begin{tabular}{|c|c|c|c|}
\hline
initial state &  final  state 
& Dynamical factor $(D)$ & Sommerfeld factor $(|A|^2)$  \\[2pt]
\hline
$\chi^0 \chi^0\ (S=0)$ & 
$W^+ W^-$ & 
${1\over 2} \left[1- x_W^2\right] \left[1- {x_W^2\over 2} \right]^{-2}$ & 
$|A^{(0,0)}_{21}+\sqrt{2} A^{(0,0)}_{22}|^2$ \\[2pt]
& $ZZ$ & 
$c_W^4 \left[1- x_Z^2\right] \left[1- {x_Z^2\over2} \right]^{-2} $ & 
$|A^{(0, 0)}_{21}|^2$ \\[2pt]
& $Z\gamma$ & 
$2 c_W^2 s_W^2$ & 
$|A^{(0, 0)}_{21}|^2$\\[2pt]
& $\gamma\gamma$ & 
$ s_W^4  $ & 
$|A^{(0, 0)}_{21}|^2$ \\[2pt]
\hline
$\chi^+ \chi^-\ (S=0)$ & 
$W^+ W^-$ & 
$ \frac{1}{2}\left[1- x_W^2\right] 
\left[1- {x_W^2\over 2} \right]^{-2}$ & 
$|A^{(0,0)}_{11}+\sqrt{2} A^{(0,0)}_{12}|^2$ \\[2pt]
& 
$Z Z$ & 
$c_W^4 \left[1- x_Z^2\right]
\left[1- {x_Z^2\over 2} \right]^{-2}$ & 
$|A^{(0,0)}_{11}|^2$\\[2pt]
& 
$Z \gamma$ & 
$2 c_W^2 s_W^2  $ & 
$|A^{(0,0)}_{11}|^2$\\[2pt]
& 
$\gamma \gamma$ & 
$s_W^4 $ & 
$|A^{(0,0)}_{11}|^2$\\[2pt]
\hline
$\chi^+ \chi^-\ (S=1)$ & 
$W^+ W^-$ & 
$\frac{1}{48} f_1(x_W, x_Z)$ & 
$|A^{(0,1)}|^2$ \\[2pt]
& 
$Z h$ & 
$\frac{1}{48} f_3(x_Z, x_h)$ & 
$|A^{(0,1)}|^2$ \\[2pt]
& 
$f \bar{f}$ & 
$\frac{N_C}{6}(T^{f}_{3L})^2 
f_4(x_Z, x_f) $ & 
$|A^{(0,1)}|^2$ \\[2pt]
\hline
$\chi^+ \chi^0\ (S=0)$ & 
$W^+ Z$ & 
$\frac{c_W^2}{2}  f_6(x_W, x_Z)$&
$|A^{(1,0)}|^2$ \\[2pt]
& 
$W^+ \gamma$ & 
$\frac{s_W^2}{2}$ & 
$|A^{(1,0)}|^2$ \\[2pt]
\hline
$\chi^+ \chi^0\ (S=1)$ & 
$W^+ Z$ & 
$\frac{1}{48} f_2(x_W,x_Z)  $ & 
$|A^{(1,1)}|^2$ \\[2pt]
& 
$W^+ h$ &
$\frac{1}{48} f_3(x_W,x_h)$ & 
$|A^{(1,1)}|^2$ \\[2pt]
& 
$u \bar{d}$ & 
$\frac{1}{4}
f_5(x_W, x_d, x_u) $ & 
$|A^{(1,1)}|^2$ \\[2pt]
&
$\nu \bar{e}$ &
$\frac{1}{12}
f_5(x_W, x_e, 0) $ & 
$|A^{(1,1)}|^2$ \\[2pt]
\hline
$\chi^+ \chi^+\ (S=0)$ & 
$W^+ W^+$ & 
${1\over 2}  \left[1- x_W^2\right]\left[1- {x_W^2\over 2} \right]^{-2}$ & 
$|A^{(2,0)}|^2$ \\[2pt]
\hline
\end{tabular}
\caption{Table of parameters to calculate annihilation cross sections
  for each initial and final states.  Tree-level cross sections are
  recovered by setting
  $A^{(0,0)}_{11}=A^{(0,0)}_{22}=A^{(0,1)}=A^{(1,0)}=A^{(1,1)}=A^{(2,0)}=1$
  and $A^{(0,0)}_{21}=A^{(0,0)}_{12}=0$.  Here, $x_I$ is the mass of
  the particle $I$ normalized by $m_{{\tilde W}}$, $x_I\equiv
  m_I/m_{{\tilde W}}$.  $f$ denotes quarks and leptons.  $N_C$ is the
  color factor; $N_C=3$ ($1$) for quarks (leptons).  For the
  calculation of the Sommerfeld factors $A^{(Q,S)}$ for states with
  charge $Q$ and spin $S$, see \cite{Hisano:2006nn}.  }
\label{Tab:cs}
\end{center}
\end{table}%

Here we present the possible decay modes and the decay widths for each
set of $(Q, S)$, which are summarized in Table \ref{Tab:cs}.  The
annihilation cross section to the $f \, f'$ final state is obtained by
\begin{eqnarray}
  \sigma v = c
  \frac{\pi^2 \alpha_2^2}{m_{\tilde W}^2}
  \sum_{S=0,1} (2S+1) P(x_f, x_{f'}) D |A|^2,
\end{eqnarray}
where $c=2$ for the annihilation of identical particles, otherwise
$c=1$.
The dynamical factor $D$ and the Sommerfeld factor $|A|^2$ are given
  in Table \ref{Tab:cs}.

The phase factor $P(x_1, x_2)$ is given by
\begin{eqnarray*}
  P(x_1,x_2) &=& 
  \sqrt{1-{x_1^2+x^2_2 \over 2} + {(x_1^2-x_2^2)^2 \over 16}}.
\end{eqnarray*}
In addition, the functions $f_i(x)$ are defined as
\begin{eqnarray*}
  f_1(x_W,x_Z) &=& \frac{4(1-x_W^2)(4+20x_W^2+3x_W^4)(2+x_W^2-x_Z^2)^2 }
  {(2-x_W^2)^2(4-x_Z^2)^2},
  \nonumber\\
  f_2(x_W,x_Z) &=&
  \frac{\left\{ (4-x_W^2-x_Z^2)^2-4 x_W^2x_Z^2 \right\}
    (16+40x_W^2+40x_Z^2+10x_W^2x_Z^2+x_W^4+x_Z^4)}
  {(4-x_W^2)^2(4-x_W^2-x_Z^2)^2},
  \nonumber\\
  f_3(x,x_h) &=&
  \left(1-\frac{x^2}{4} \right)^{-2}
  \left(1+\frac{5x^2-x_h^2}{2}+\frac{(x^2-x_h^2)^2}{16}\right),
  \nonumber\\
  f_4(x_Z, x_f) &=&
  \left(1- {x_Z^2\over 4}\right)^{-2}
  \left[
  \left( 1 -  c_f x_Z^2 + {c_f^2 \over 2} x_Z^4 \right) 
  -
  \frac{x_f^2}{4}
  \left(1+ 2 c_f x_Z^2 - c_f^2 x_Z^4 \right) 
  \right],
  \nonumber\\
  f_5(x_W, x_d, x_u) &=& 
  \left(1- {x_W^2\over 4}\right)^{-2}
  \left(1- {x_u^2 + x_d^2 \over 8} - { (x_u^2-x_d^2)^2 \over 32}\right),
  \nonumber\\
  f_6(x_W, x_Z) &=& 
  \left(1-{x_W^2+x_Z^2 \over 4} \right)^{-2}
  \left(1- {x_W^2+x_Z^2\over 2} + {(x_W^2-x_Z^2)^2 \over 16} \right),
\end{eqnarray*}
where $c_f={Q_f s_W^2 / (2 T^f_{3L})}$ for the particle with the
electric charge $Q_f$ and the weak isospin $T^f_{3L}$.

\end{document}